\documentclass[twoside]{dis09}
\usepackage[latin1]{inputenc}
\usepackage[dvips]{graphicx,epsfig,color}
\usepackage{wrapfig,rotating}
\usepackage{amssymb,amsmath,array}

\newcommand{\com}[1]{{{#1}}} 
 
\begin{document}
\title {Strange and Anti-strange Sea Distributions
from $\boldmath{\nu N}$ Deep Inelastic Scattering}

\author{S. Alekhin$^1$, S. Kulagin$^2$, and R. Petti$^3$
%
%
\vspace{.3cm}\\
%
1- Institute for High Energy Physics \\
142281 Protvino, Moscow Region - Russia\\
%
\vspace{.1cm}\\
2- Institute for Nuclear Research of the Academy of Sciences of Russia \\
117312 Moscow - Russia\\
\vspace{.1cm}\\
3- Department of Physics and Astronomy, University of South Carolina \\
Columbia SC 29208 - USA\\
}

\maketitle

\begin{abstract}
{\small
We perform QCD fit of the nucleon strange and anti-strange sea distributions to 
the neutrino and anti-neutrino  
dimuon data by the CCFR and NuTeV collaborations, supplemented 
by the inclusive charged lepton-nucleon 
Deep Inelastic Scattering (DIS) and Drell-Yan data. 
The effective semi-leptonic charmed-hadron branching ratio
is constrained from the inclusive charmed 
hadron measurements performed 
by the FNAL-E531 and CHORUS neutrino emulsion experiments 
as $B_\mu=(8.8\pm0.5)\%$. 
We obtain a strange sea suppression factor 
{$\kappa(20~{\rm GeV}^2)=0.62\pm0.04({\rm exp.})\pm0.03({\rm QCD})$}.  
An $x$-distribution of total strange sea obtained in the fit is
slightly softer than the non-strange sea, and an asymmetry 
between strange and anti-strange
quark distributions is consistent with zero (integrated over $x$ it is equal to
{$0.0013\pm 0.0009({\rm exp.}) \pm 0.0002({\rm QCD})$ 
at the scale of $20~{\rm GeV^2}$}).
}
\end{abstract}

At Bjorken variable $x\lesssim 0.2$ the strange 
quarks give 
\com{important} 
contribution to the total quark sea 
in the nucleon. Therefore accurate determination of the strange sea is 
necessary for interpretation of the precise hadron-collider and fixed-target
experimental data. In particular, 
\com{%
the presence of a small positive $s-\bar s$ asymmetry in the nucleon may explain \cite{Davidson:2001ji} recent anomalous result
on the weak mixing angle in the NuTeV experiment \cite{nutevanom}.
}
The best constraint on 
the strange sea comes from the DIS  
neutrino-nucleon dimuon production. This process  
stems from the charged-current production of the charm quark decaying 
semileptonically with a secondary muon in the final state. 
\com{%
This mechanism is particularly sensitive to the strange quark distributions,
as the contribution from non-strange quarks are greatly 
suppressed due to smallness of the corresponding 
Cabibbo-Kobayashi-Maskawa (CKM) matrix elements.
}
This allows to disentangle 
the strange distribution from other quarks in the global PDF fit, which 
includes the dimuon data. 

\begin{figure}
\centerline{\includegraphics[width=\columnwidth,height=7cm]
{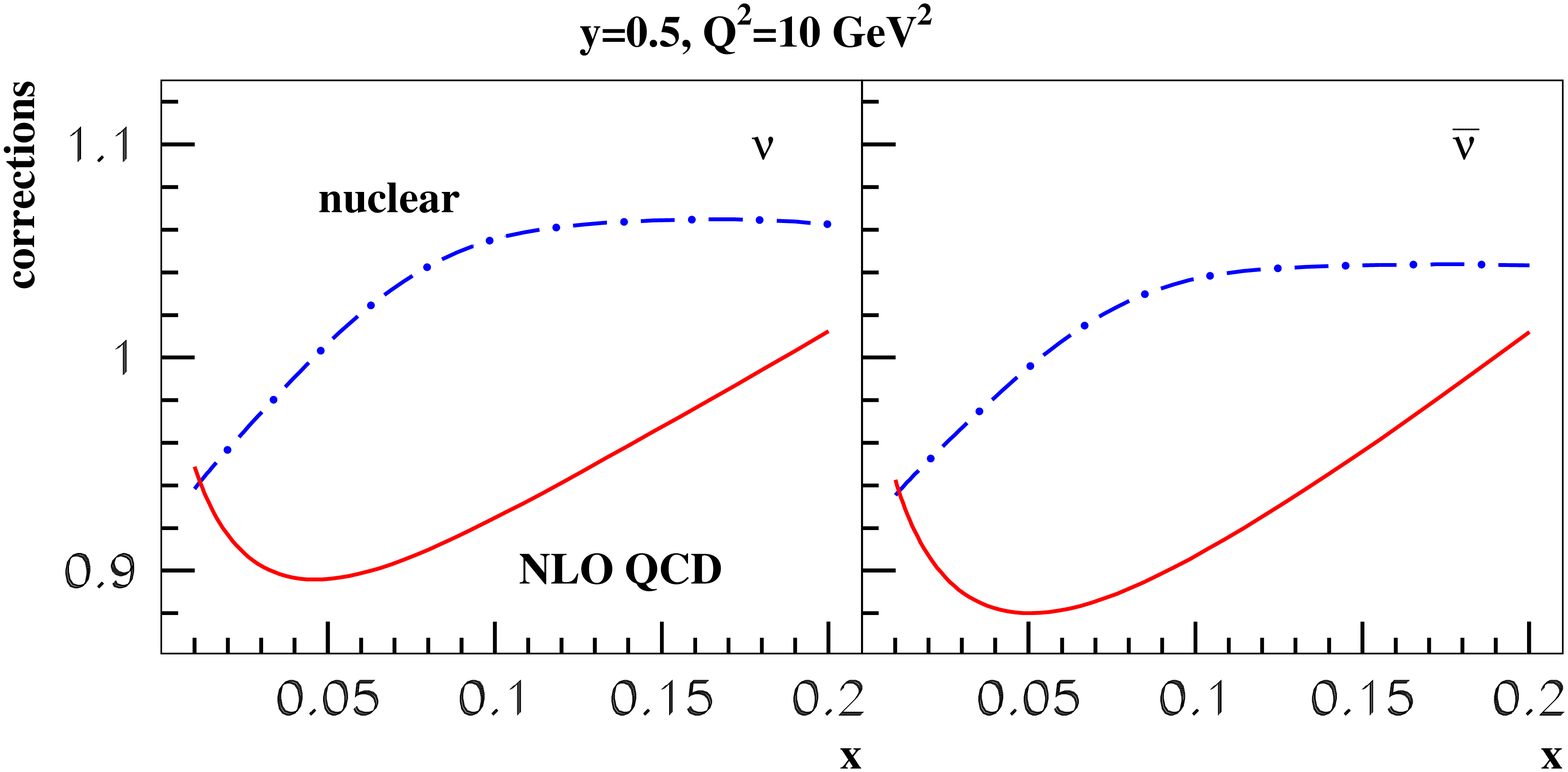}}
\caption{\small The values of the corrections 
on the calculated dimuon cross sections employed in our analysis 
(solid lines: the NLO QCD corrections, dotted-dashes: the nuclear 
correction for the case of iron target) for the neutrino (left panel) and 
anti-neutrino (right panel) beam at the representative kinematics 
of the NuTeV and CCFR dimuon data, $Q^2=10~{\rm GeV}^2,~y=0.5$.
}
\label{fig:corr}
\end{figure}
We extract the strange distribution from the dimuon data by the 
CCFR and NuTeV 
experiments~\cite{Bazarko:1994tt,Goncharov:2001qe,Mason:2007zz}.
Those data were collected in the runs with the neutrino and anti-neutrino 
beams and 
allow separate determination of the strange and anti-strange distributions
at $0.01\lesssim x \lesssim 0.3$ and the values of the 
momentum transfer $1~{\rm GeV}^2 \lesssim Q^2\lesssim 200~{\rm GeV}^2$.
\com{In order to consistently account for} 
the contribution from other quarks and gluons,
the dimuon data are supplemented by the inclusive DIS data and the fixed-target 
Drell-Yan data
(cf.  Ref.~\cite{Alekhin:2006zm} for details of the data selection).
The analysis is performed with account of the NNLO correction to 
the parton evolution~\cite{Moch:2004pa} 
and the massless coefficient functions~\cite{Zijlstra:1992qd}. The 
heavy-quark DIS production  
is calculated in the 3-flavour factorization scheme with the 
\com{perturbative}
corrections up to 
the NLO~\cite{Gottschalk:1980rv}.
The latter 
are illustrated in Figure~\ref{fig:corr} for the representative kinematics 
of the data used in the fit and the factorization scale 
$\mu=\sqrt{Q^2+m_c^2}$, where $m_c$ is the charm quark mass. 
At small $x$ the NLO corrections 
of Ref.~\cite{Gottschalk:1980rv} reduce the value of calculated 
cross sections therefore they increase the strange distributions extracted 
from the data. Corrections for nuclear effects in the 
neutrino-nucleon DIS, which include the Fermi motion and binding, 
neutron excess and shadowing, nuclear pion excess and the off-shell 
effects~\cite{KP04} are also taken 
into account in our analysis. The value 
of nuclear corrections depends on the beam type  
(cf. Figure~\ref{fig:corr}), therefore they affect the charge asymmetry in 
the strange distribution $x[s(x)-\bar{s}(x)]$. 
Other details of the analysis including treatment of the data and their 
uncertainties can be found elsewhere~\cite{alekhin1}.

\begin{figure}
\centerline{\includegraphics[width=\columnwidth,height=6cm]
{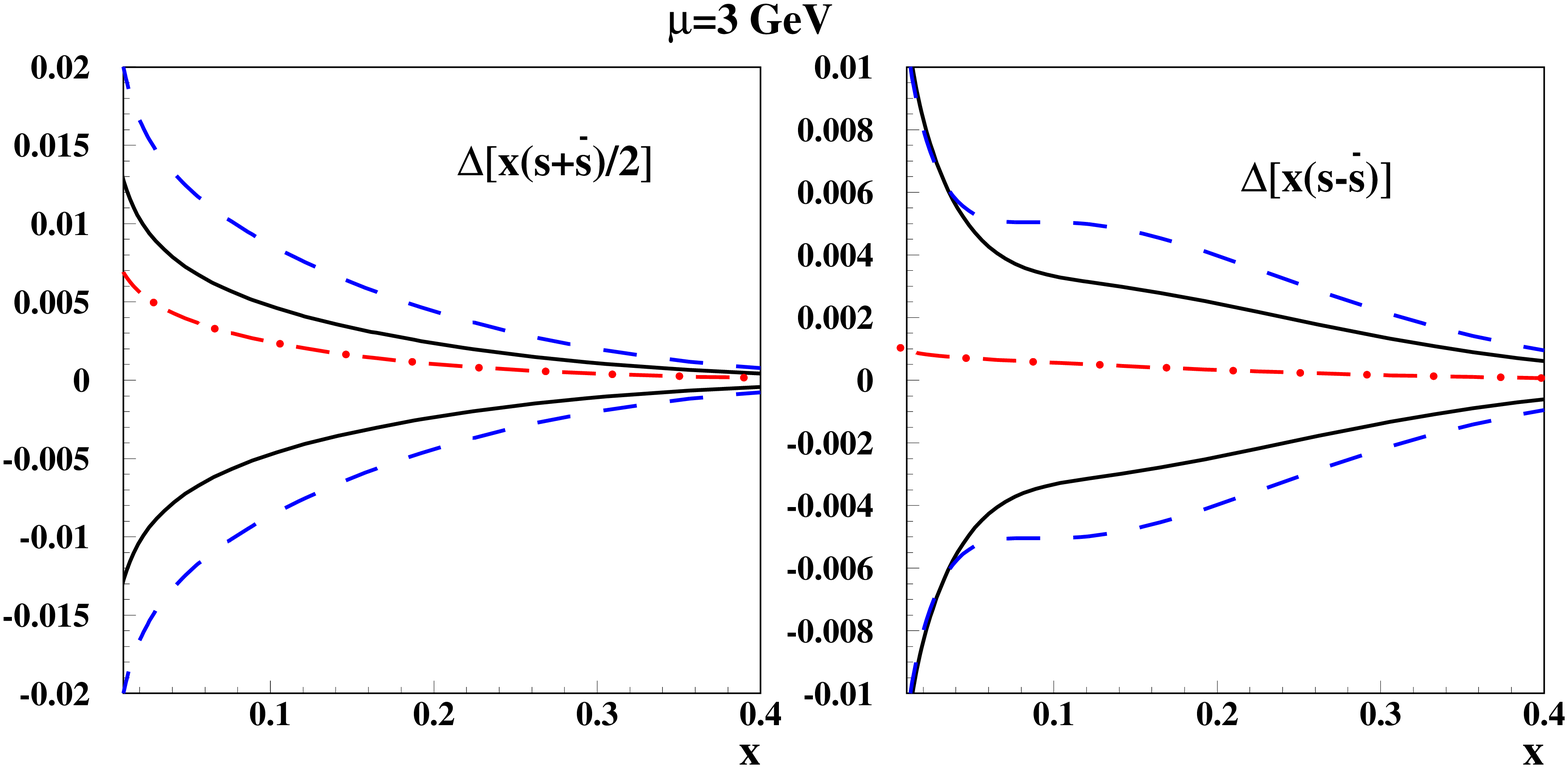}}
\caption{\small The $\pm 1\sigma$
uncertainties in the $C$-even (left panel) and 
$C$-odd (right panel) combination of the 
strange sea distributions determined in our fit with (solid lines) 
and without (dashed lines) the constraint on $B_\mu$ at the value of 
factorization scale $\mu=3~{\rm GeV}$. The
difference between distributions obtained in these two variants 
of the fit is shown by the dashed-dotted lines.
}
\label{fig:unc}
\end{figure}
\com{Assuming universality of the charmed hadron fragmentation function,}
the dimuon production cross section is related to the charm production 
cross section as follows
\begin{equation}
\frac{d\sigma_{\mu\mu}(E_\mu > E_\mu^0)}{dxdy}=
\eta_\mu B_\mu
\frac{d\sigma_{\rm charm}}{dxdy},
\label{eqn:cut}
\end{equation}
where $y$ is inelastisity, $\eta_\mu$ is the acceptance correction 
accounting for the cut 
$E_\mu > E_\mu^0$  imposed on the muon energy to reject background, 
and $B_\mu$ is the effective semileptonic charm quark branching ratio.
The value of $B_\mu=\sum f_{\rm h} Br(h\to\mu X)$, where $f_{\rm h}$
and $Br(h\to\mu X)$ are the rate and
the semileptonic branching ratio of the charmed hadron $h$ produced in the 
(anti)neutrino-nucleon scattering, respectively, 
$h=D^0,D^+,D_s^+,\Lambda_{\rm c}^+$.
The value of $B_\mu$ depends on the beam type and energy, due to $f_{\rm h}$.
For kinematics of the NuTeV and CCFR experiments
a value of $B_\mu=(9.19\pm0.94)\%$ averaged over neutrino and antineutrino
beams was obtained in Ref.~\cite{Bolton:1997pq} using the 
charmed hadron fractions measured 
by the Fermilab-E-531 experiment. 
In our analysis we fit the value of $B_\mu$ simultaneously with the PDF 
parameters. In this case
 the strange \com{quark} distributions are constrained by the 
$Q^2$-slope of the dimuon cross sections, which depend 
on the strange sea magnitude due to the PDF evolution, and the value 
of $B_\mu$ is constrained by the magnitude of the cross sections. 
This approach \com{provides a} consistent determination of $B_\mu$ and allows to 
check the beam- and energy-dependence of $B_\mu$  
for the kinematics of experiments used in the fit.
\small
\begin{wraptable}{l}{0.7\columnwidth}
\centerline{\begin{tabular}{l|cc|cc} \hline
Measurement    & \multicolumn{2}{c|}{$E_\nu > 5$ GeV} & \multicolumn{2}{c}{$E_\nu > 30$ GeV} \\ \hline\hline   
E531 charmed fractions~\cite{Bolton:1997pq} && $7.86\pm0.49$ && $8.86\pm0.57$  \\ 
CHORUS direct~\cite{KayisTopaksu:2005je} && $7.30\pm0.82$ && $8.50\pm1.08$  \\ 
CHORUS charmed fractions~\cite{DiCapua08} && $9.11\pm0.93$ &&   \\ \hline
Weighted average && $7.94\pm0.38$ && $8.78\pm0.50$  \\ \hline\hline 
\end{tabular}}
\caption{ \small  
Semileptonic branching ratio $B_\mu(\%)$ from direct measurements in the E531 and CHORUS 
emulsion experiments. The last row corresponds to our weighted average. 
\label{tab:Bmu}
}
\end{wraptable}
\normalsize
We do not observe statistically significant 
beam- and energy-dependence of $B_\mu$. The averaged over beam type and energy 
value of \com{$B_\mu=(9.1\pm1.0)\,\%$} obtained in our fit is 
in good agreement with the result of Ref.~\cite{Bolton:1997pq}. 
The value of $B_\mu$ is anti-correlated with the magnitude of the 
strange sea, therefore an additional constraint on $B_\mu$ can improve 
the strange sea determination. 
The value of $B_\mu$ obtained in Ref.~\cite{Bolton:1997pq} does not provide 
efficient constraint since its uncertainty
\com{is consistent with the corresponding uncertainty in our fit.} 
\com{The} uncertainty in $B_\mu$
can be \com{reduced} 
\com{by utilizing the results of the}
recent measurement of the charmed hadron fractions by the 
CHORUS experiment~\cite{KayisTopaksu:2005je,DiCapua08}
\com{as well as the updated} 
determination of the semileptonic branching ratios 
Ref.~\cite{PDG08}.
The value of $B_\mu$ obtained with these inputs is given 
in Table~1. The uncertainty \com{of this} determination is \com{about factor of 2}  smaller than 
\com{that}
of Ref.~\cite{Bolton:1997pq} and the central value \com{of $B_\mu$} is somewhat lower
(cf. Table~1.). In the variant of our fit 
with the constraint of Table~1 \com{for the beam energy cut} 
$E_\nu>30~{\rm GeV}$,
the central values of the strange and antistrange 
distributions are smaller, 
correspondingly. In this fit the uncertainty in the 
$C$-even combination $x[s(x)+\bar{s}(x)]/2$
is about factor of 2 smaller than for the fit with no constraint on $B_\mu$
(cf. Figure~\ref{fig:unc}). For the $C$-odd combination $x[s(x)-\bar{s}(x)]$\com{, the}
effect of the \com{constrained $B_\mu$ is weaker because of a partial cancellation in the}
difference. 

\begin{figure}
\centerline{\includegraphics[width=\columnwidth,height=6cm]
{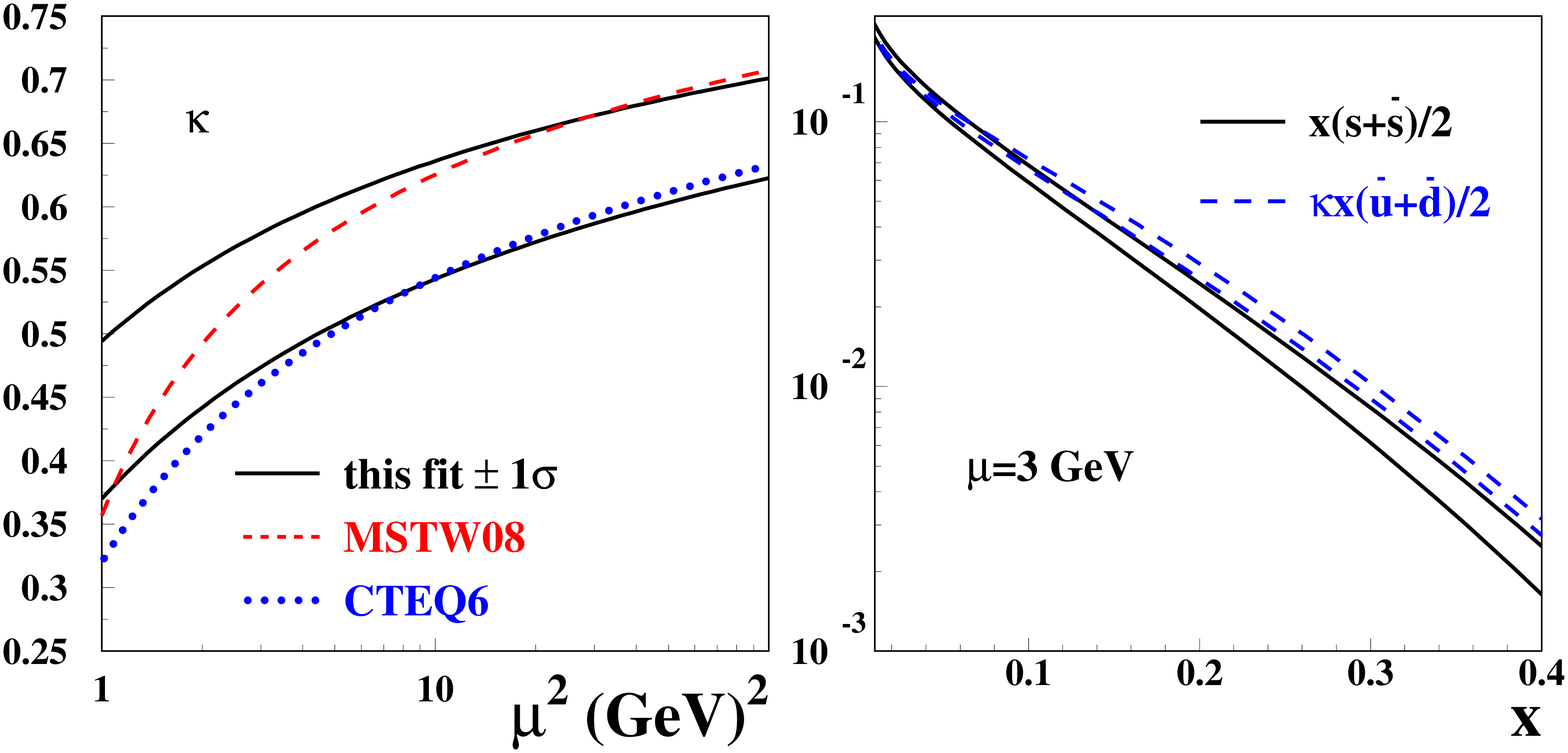}}
\caption{\small Left panel: The $\pm 1\sigma$ band for the strange sea 
suppression factor $\kappa$ with respect to the factorization scale $\mu$
as obtained in our analysis (solid lines) in comparison with 
ones for the MSTW08 (dashes) and CTEQ6 (dots) PDF sets. 
Right panel: The $\pm 1\sigma$ band for the $C$-even combination of the 
strange sea distributions determined in our fit (solid lines)
compared to the non-strange one scaled by $\kappa$ (dashes) at 
$\mu=3~\rm{GeV}$.
}
\label{fig:ssup}
\end{figure}

In the fit with the \com{constrained  $B_\mu$,}
we obtain the values of $\chi^2/NDP$ of 63/89 and 38/89 for the 
CCFR and NuTeV data sets, respectively. This is statistically consistent with 
the effective number of degrees of freedom for these experiments
introduced in Ref.~\cite{Goncharov:2001qe}, \com{in order} to take into account 
statistical correlation between the different data points that is roughly
twice smaller than $NDP$. The value of the strange sea suppression 
factor $\kappa$, which is the ratio of sum of momentum carried by the strange and anti-strange 
quarks to the one carried by the sea up- and down-quarks, is given in 
Figure~\ref{fig:ssup}. Due to the QCD evolution the value of $\kappa$ rises 
with the factorization scale $\mu$. The value of 
$\kappa(20~{\rm GeV}^2)=0.48\genfrac{}{}{0cm}{1}{+0.06}{-0.05}$ 
was obtained in the NLO fit of Ref.~\cite{Bazarko:1994tt} based on the 
CCFR dimuon data. In our fit we \com{get a} bigger value, 
$\kappa(20~{\rm GeV}^2)=0.62\pm0.04$. 
This \com{is} because the non-strange sea distribution used in
Ref.~\cite{Bazarko:1994tt} is 
\com{not consistent with the distribution of the present fit as well as with}
other modern PDF sets~\cite{Martin:2009iq,Lai:2007dq}.
Our value of $\kappa$ is somewhat smaller
than the value for the MSTW08 PDFs set of Ref.~\cite{Martin:2009iq} 
and is somewhat  bigger than one for the CTEQ6 PDFs set of Ref.~\cite{Lai:2007dq}. 
In both cases, however, the 
discrepancy is within uncertainty in our determination of $\kappa$,
which is obtained 
from propagation of the statistical and systematical errors in the data with 
\com{the account of correlations.}
\com{As can be seen from Figure~\ref{fig:ssup}, 
the $C$-even strange sea is somewhat steeper than the non-strange sea and at small $x$ overshoots the latter.} 
At $x\lesssim 0.2$ the strange sea is enhanced 
due to the NLO corrections to the charm production coefficient functions. 
\com{Were} these corrections \com{dropped,}
we obtain the value of $\kappa(20~{\rm GeV}^2)=0.55\pm0.13$.
The uncertainty in $\kappa(20~{\rm GeV}^2)$ due to the higher-order QCD 
corrections, estimated as the change due to variation of the 
dimuon cross section factorization scale from $\sqrt{Q^2+m_c^2}$
to $Q$, is 0.03. This is comparable to the experimental uncertainty 
in $\kappa$. In this sense the theoretical and experimental uncertainties in 
our determination of $\kappa$ are balanced, and, in order to \com{further} improve 
it, more experimental data are needed and 
the NNLO QCD corrections have to be taken into account. 

In a variant of our fit with 
only the NuTeV dimuon data, the strangeness charge asymmetry 
is somewhat positive. This is consistent with the results of the NLO 
analysis of Ref.~\cite{Mason:2007zz}. The CCFR data prefer 
somewhat negative asymmetry and the result averaged over these two experiments
is consistent with 0 in a wide range of $\mu$, even \com{despite} it 
rises with $\mu$ due to the NNLO corrections to the PDF 
evolution~\cite{Catani:2004nc}.  Furthermore, 
if we fix the strange sea asymmetry 
at 0, the value of $\chi^2$ rises by \com{about 1 unit} only. 
For the fit based on the combined 
NuTeV and CCFR dimuon data we obtain 
{$S^-=\int^1_0x[s(x)-\overline{s}(x)]dx=0.0013\pm 0.0009$} 
at $\mu^2=20~{\rm GeV}^2$, that is also consistent with 0 within the 
uncertainties. The theoretical uncertainty in $S^-$ 
due to the factorization scale variation is 
0.0002, much smaller than \com{both} the experimental uncertainty and 
the theoretical uncertainty in $\kappa$.

In summary, we obtain the strange sea suppression factor 
{$\kappa(20~{\rm GeV}^2)=0.62\pm0.04({\rm exp.})\pm0.03({\rm QCD})$}, 
the most precise value available.
The $x$-distribution of \com{the} total strange sea is slightly softer 
than the non-strange one, and 
the integral strange sea charge asymmetry  
$S^-(20~{\rm GeV}^2)=0.0013\pm 0.0009({\rm exp.}) \pm 0.0002({\rm QCD})$ is 
consistent with 0 within uncertainties. 

{\bf Acknowledgments.}
This work \com{was} partially supported by the \com{Russian Foundation for Basic Research, grant} 06-02-16659. 
R.P. thanks USC for supporting this research.

\begin{footnotesize}

\end{footnotesize}

\end{document}